\documentclass[prd,showpacs]{revtex4}

\usepackage{graphicx}

\newcommand{\ieps}[1]{\includegraphics[height=3in]{#1}}

\begin{document}

\title{Simulations of laser locking to a LISA arm}
\author{Julien Sylvestre}
\altaffiliation[Also at: ]{LIGO Laboratory, California Institute of Technology,\\MS 18-34, Pasadena, CA 91125}
\email{jsylvest@ligo.caltech.edu}
\affiliation{Jet Propulsion Laboratory, California Institute of Technology, Pasadena, California 91109}

\date{\today}

\begin{abstract}
We present detailed numerical simulations of a laser phase stabilization scheme for LISA, where both lasers emitting along one arm are locked to each other. Including the standard secondary noises and spacecraft motions that approximately mimic LISA's orbit, we verify that very stable laser phases can be obtained, and that time delay interferometry can be used to remove the laser phase noise from measurements of gravitational wave strains. Most importantly, we show that this locking scheme can provide significant simplifications over LISA's baseline design in the implementation of time delay interferometry.
\end{abstract}


\maketitle

\section{Introduction}
The LISA mission consists in three spacecraft (S/C) to be launched over the next decade in order to implement laser interferometry in space for the observation of gravitational waves of galactic, extra-galactic, and potentially cosmological origin.
The S/C will be placed on heliocentric orbit in a triangular configuration,  where each arm of the triangle will be approximately $5 \times 10^9$ m long.
Gravitational waves will be detected by monitoring the relative position of the S/C with high accuracy, leading to a gravitational wave strain sensitivity of $\sim 10^{-23}/\sqrt{\rm Hz}$ from 0.1 mHz to 0.1 Hz.

One of the most significant technical challenge to achieve this level of accuracy is the cancellation of the laser phase noise.
A number of signal processing techniques, known as Time Delay Interferometry (TDI), have been developed in order to synthesize noise-cancelling interferometers from the various phase measurements that LISA will provide \cite{TDI1,TDI2,TDI3,TDI4,TDI5,TDI6,TDI7}.
While these techniques will certainly be very useful, they impose a number of relatively strict requirements on the system \cite{TDIImplementation}, such as a good knowledge of the arm lengths at all time, precise synchronization of the clocks between S/C, etc.

An intriguing idea was recently proposed in Ref. \cite{SGMS} to significantly reduce the laser phase noise: the very stable arms of the LISA constellation could be used as frequency references to lock the lasers.
We describe the control loop topology proposed by \cite{SGMS} (referred to as ``laser self-locking'') in section \ref{LockingScheme}, where we also present a time domain analysis of the behavior of the transients introduced by the locking procedure.
We then present in section \ref{NumericalSimulations} the results from a set of realistic numerical simulations of LISA with laser self-locking, where all standard secondary noises are included, and where the orbital motion of the S/C is approximately included.
We finally show in section \ref{TimeDelayInterferometry} results for the use of TDI to bring the reduced laser phase noise below the other noises.
We describe in particular how some of the technical challenges of implementing TDI in LISA's baseline design are simplified by using laser self-locking.

\section{Locking scheme} \label{LockingScheme}
\begin{figure}
\begin{center}
\ieps{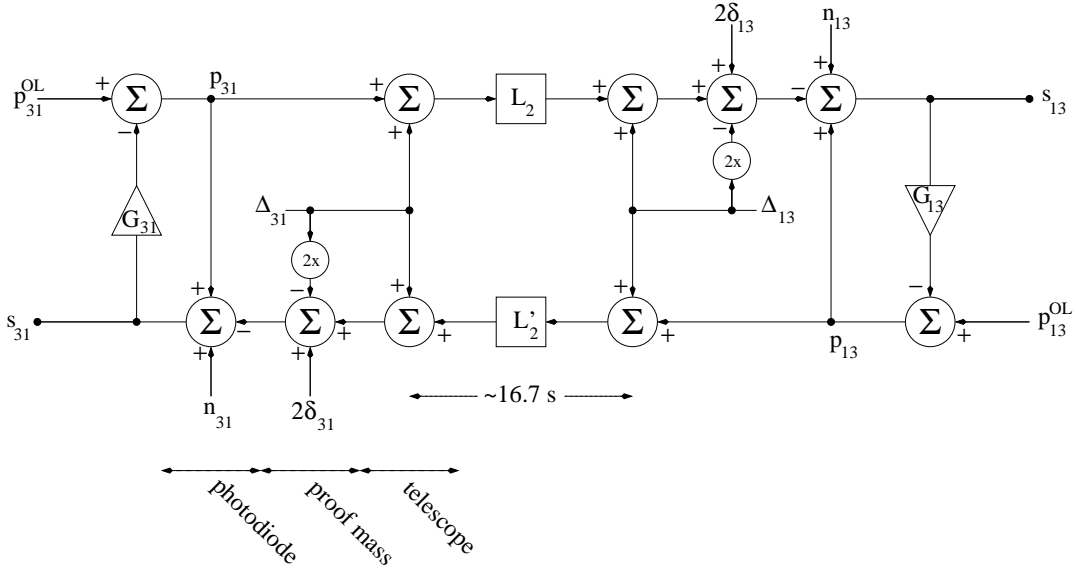}
\end{center}
\caption{Diagram of the control system. S/C 1 is left of the 16.7 delay lines representing arm 2, while S/C 3 is to its right.}
\label{fig:topology}
\end{figure}

As illustrated by Figure \ref{fig:topology}, we use the same locking scheme as \cite{SGMS}, although we add realistic acceleration and optical noises to our model.
Open loop laser noises $p^{\rm OL}_{31}$ and $p^{\rm OL}_{13}$ are generated on-board S/C 1 and 3, respectively.
They then follow a similar path.
$p^{\rm OL}_{31}$ is collimated by the telescope on S/C 1, where it acquires noise $\Delta_{31}$ from the S/C motion, before being propagated down arm 2 towards S/C 3.
Upon reception after a travel time $L_2$\cite{Rotating}, it is exposed to the noise $\Delta_{13}$ from the motion of S/C 3.
It is bounced off the proof mass (acquiring S/C noise $2\Delta_{13}$ and proof mass acceleration noise $2\delta_{13}$), and interacts with the local laser on a photodiode, where the optical noise $n_{13}$ is added to produce the electrical signal $s_{13}$. 
This signal would be the normal phase measurement in the absence of locking.
In the present configuration, however, the signal $s_{13}$ is fed back to the laser phase actuator through a controller $G_{13}$ in order to close the control loop.

The equations that relate $p_{31}$ and $p_{13}$ to the other noises and to the controllers gain are derived in \cite{SGMS}.
Although the authors of \cite{SGMS} note the presence of a strong, slowly decaying quasiperiodic signal with period equal to the round-trip light-time, they do not include it explicitly in their expressions for the noise spectra.
This fact was first pointed out by Tinto \cite{Tinto} and others.
However, we show below that the quality factor of this signal is very high, so that its power spectrum consists only in narrow peaks at harmonics of the inverse of the round-trip light-time.
Its effect on data analysis should therefore be negligible, just like high quality factor oscillations from wire resonances, for instance, have a negligible effect on the data analysis for ground-based interferometers.

As a simplification of the diagram of Fig. \ref{fig:topology}, neglect all secondary noises, and assume that S/C 3 is an ideal mirror, so that
\begin{equation}
p_{31}(t) = p^{\rm OL}_{31}(t) - G_{31}*\left[ p_{31}(t) - p_{31}(t - 2L_2) \right], \label{eq:lock}
\end{equation}
where the $*$ operator denotes convolution.
It is assumed that this equation holds for $t > 2 L_2$, with the control loops being closed at $t=0$.
For $t < 2 L_2$, is is assumed that no light is reflected off S/C 3 back to S/C 1.
Without loss of generality, the solution to Eq. \ref{eq:lock} is assumed to take the form
\begin{equation}
p_{31}(t) = f(t) Q(t) + P(t)
\end{equation}
for $t > 2 L_2$, where 
\begin{eqnarray}
Q(t) = Q(t - 2 L_2), \label{eq:periodic} \\
f(2 L_2) = 1,
\end{eqnarray}
and $P(t)$ is an arbitrary function with no power at frequencies $f = n/2L_2$, for $n = 1, 2, ...$
Substituting into Eq. \ref{eq:lock} and rearranging,
\begin{eqnarray}
f(t) Q(t) + G_{31}*\left[ f(t) Q(t) - f(t-2L_2) Q(t-2L_2) \right] \nonumber \\
+ P(t) + G_{31}*\left[ P(t) - P(t-2L_2) \right] = p^{\rm OL}_{31}(t).\label{eq:lockl}
\end{eqnarray}
In order for $P$ to not have any periodic component at frequency $1/2L_2$, the first line in Eq. \ref{eq:lockl} must vanish.
Using Eq. \ref{eq:periodic} and assuming that the convolution is dominated by the gain $g$ at frequency $1/2L_2$, we have
\begin{equation}
\left[ (1+g) f(t) - g f(t-2L_2) \right] Q(t) = 0,
\end{equation}
i.e.,
\begin{equation}
f(t) = \frac{g}{1+g} f(t-2L_2).
\end{equation}
The solution to this equation is
\begin{equation}
f(t) = e^{-t/\lambda}
\end{equation}
where
\begin{equation}
\lambda = \frac{2 L_2}{\log(1+1/g)} \simeq 2 g L_2. \label{eq:decayrate}
\end{equation}
Since $g$ is typically very large, the quasiperiodic transient decays on a large number of round-trip light-times, and therefore produces only a few narrow peaks in the noise spectrum, which have a width $\sim 1/\lambda$ at frequencies that are multiples of $1/2L_2$.

It is also straightforward to show that the function $Q$ is mostly defined by the open loop laser noise from both S/C in the time $L_2$ just after the loops are closed.
Colloquially, the lasers ``load'' the arm buffer with $\sim 16.7$ s of noise after the loops are closed.
As soon as light reaches the other S/C, the lasers will try to track the phase of the light stored in the arm, so that the initial noise effectively becomes a reference signal for future frequency measurements in the closed loop configuration.

\subsection{Controller design}
According to Fig. \ref{fig:topology}, we assume that the electronic signal out of the phasemeter on every S/C is filtered by a high gain digital controller, and is fed back to the laser using a phase actuator, which frequency response is described by a pole at DC (as in \cite{SGMS}).
This DC pole provides a $1/f$ attenuation, and a phase shift of $-\pi/2$ radian at all frequencies.
In addition, the time delay for the propagation of the light down the arms provides a transfer function with a significant attenuation and a $+\pi/2$ phase shift at low frequencies. 
Its amplitude goes to zero at multiples of the frequency defined by the inverse of the round-trip light-time, where the phase also drops to $-\pi/2$.

In order to have a controller that is stable at all times during the ramping up of its gain, the total phase of the phase actuator, time delay, and controller transfer functions must be larger than $-\pi$ at all frequencies by some phase margin $\phi$.
We chose a conservative $\phi = 0.35$ rad. Well below the first round-trip light-time resonance at 0.03 Hz, the phase shift from the delay line cancels the phase shift from the phase actuator, so that $\pi - \phi$ is available to the controller, and a steep roll-off of $f^{-0.8}$ is possible ($f^{-1.8}$ with the actuator DC pole).
At its first null, the phase of the time delay transfer function is $-\pi/2$, so we have to provide $+\phi$ of phase compensation at 30 mHz. 
Given the $f^{-1}$ response from the actuator, the gain then goes as $f^{-0.8}$.
At higher frequencies, the phase shift resonances are closely spaced, so we must also provide $+\phi$ of phase compensation at all frequencies and maintain the $f^{-0.8}$ attenuation.

\begin{figure}
\begin{center}
\ieps{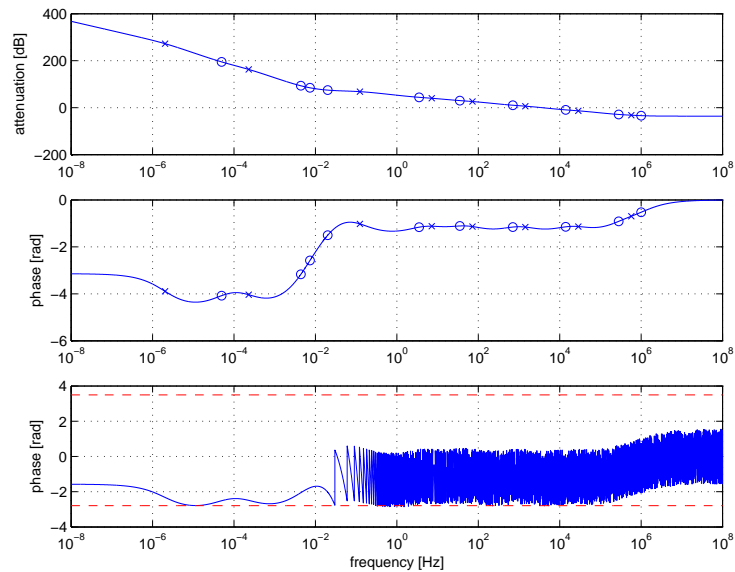}
\end{center}
\caption{Characteristics of the controller. Top: amplitude response of the controller and phase actuator (DC pole) vs. frequency. Middle: phase response of the controller and phase actuator. Bottom: phase response of the control loop, including the controller, 33.4 s delay line, and the actuator. The two horizontal dashed lines are at $\pm\pi\mp 0.35 $ rad. Circles and crosses represent the frequency location of zeros and poles, respectively.}
\label{fig:bode}
\end{figure}

This design was used to construct the controller with response shown in Fig. \ref{fig:bode} (note that the top two panels in Fig. \ref{fig:bode} include the effect of the DC pole of the phase actuator).
The gain was scaled to provide a unity gain frequency around 10 kHz, so that this controller provides close to 180 dB of attenuation at 0.1 mHz.
Although a bit more gain might be obtained by better fitting the optimal phase response ($-\pi + \phi$) around 10 mHz and below, this controller is a fairly aggressive design that allows the investigation of the limits of the self-locking procedure.

\section{Numerical Simulations} \label{NumericalSimulations}
An optimized code was used to simulate the system with realistic parameters.
At a sampling frequency of 100 kHz, open loop laser noise with an approximate $f^{-2}$ power spectrum was generated by summing successive data from a white noise series.
At the first time step, each open loop laser noise datum was propagated through the diagram of figure \ref{fig:topology} in order to calculate $s_{31}$ and $s_{13}$, as if the feedback through the controller was not present.
At the second time step, the values of $s_{31}$ and $s_{13}$ obtained from the first time step, and filtered by the controller response $G_{13}$ or $G_{31}$, were also added to the open loop laser noise.
This procedure was then repeated for all other time steps, effectively producing an integration of the system through the equivalent of the Euler scheme.

The proof mass acceleration noise ($\delta_{ij}$) was an approximate $f^{-2}$ noise generated like the open loop laser noise, while the optical path noise ($n_{ij}$)  and the spacecraft acceleration noise ($\Delta_{ij}$) approximately had $f^2$ power spectra, and were generated by taking the difference of successive data in a white noise series. 
At 1 mHz, the ratio of the rms values $p^{\rm OL}_{ij}:\Delta_{ij}:\delta_{ij}:n_{ij}$ was $1:4.88\cdot 10^{-10}:1.58\cdot 10^{-11}:3.95\cdot 10^{-12}$.

A strong anti-aliasing low-pass filter with corner frequency at 1 Hz was applied to the $f^2$ noises in order to avoid the aliasing of high frequency noise back to lower frequencies when the arms lengths were changing.
For an arm length rate of change $\dot{L} > 0$, a given noise perturbation at frequency $f$ is Doppler shifted to a different frequency $f' = f(1+\dot{L}/c)$ for every arm round-trip.
Over hundreds or thousands of round-trips, a high frequency perturbation can easily be pushed above the Nyquist frequency, and be aliased back to contaminate the measurement spectrum.

The propagation delay down the arm was simulated using a data buffer with varying length.
At any given time, the transmitted value was taken from the linear interpolation between the two nearest points in time.
The interpolation was critical in order to prevent the introduction of numerical artifacts in the calculations.
Ref. \cite{Folkner} shows that the motion of the S/C can be modeled by arm lengths that are varying with time in a sinusoidal manner.
The peak rates of change of two of the three arms are approximately equal ($\sim 1$ m/s), and are significantly less than that of the third arm ($\sim 13$ m/s), while the period of the variation of the third arm length is longer that the period of the other two arms by a factor of $\sim 3$ (1 year vs. 4 months).
As a rough approximation of these orbits, we assume a constant rate of change that is equal to 1 m/s for arm 1, and 10 m/s for arm 2.
All arm-lengths are assumed to be 16.7 s at the beginning of the simulations.

\subsection{Results}

\begin{figure}
\begin{center}
\ieps{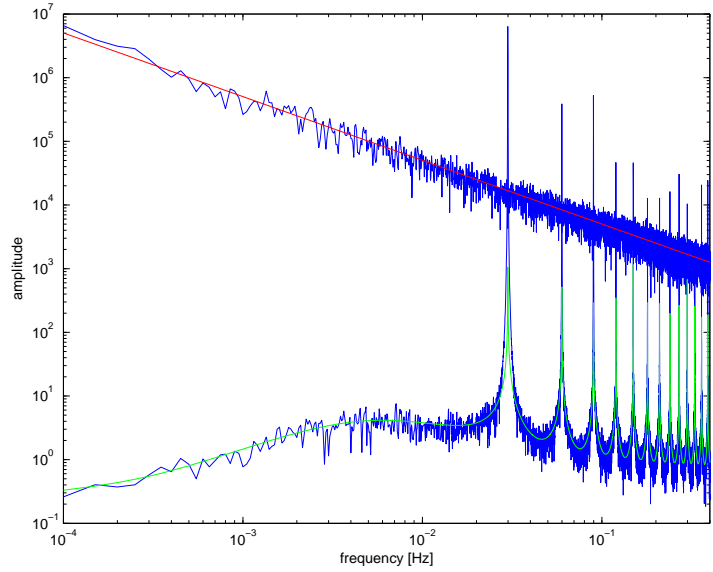}
\end{center}
\caption{(Color online) The open (top lines) and closed (bottom lines) loop laser noise amplitude spectra (arbitrary units). The upper straight (red) line is the theoretical model for the $f^{-2}$ noise, while the lower continuous (green) curve is the model for the closed loop laser noise. The spectra were constructed from $7\cdot 10^4$ s of data sampled at 100 Hz.}
\label{fig:psd}
\end{figure}

\begin{figure}
\begin{center}
\ieps{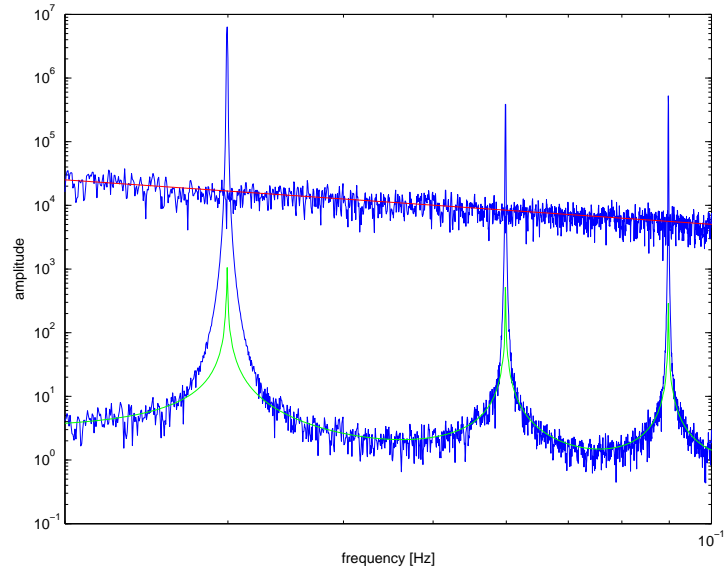}
\end{center}
\caption{(Color online) Close-up of figure \ref{fig:psd} for $0.02 {\rm Hz} < f < 0.1 {\rm Hz}$.}
\label{fig:psd_zoom}
\end{figure}

Figure \ref{fig:psd} shows the amplitude spectra of the laser noise (closed and open loop) for the 10 m/s arm change, together with the theoretical expectations for a $f^{-2}$ noise and for the closed loop noise.
The closed loop model is constructed from \cite[Eq. (12)]{SGMS}, using the known response of our controller.
The fit is excellent, except in the vicinity of the resonances at multiples of the inverse of the round-trip light-time (Fig. \ref{fig:psd_zoom}).
At those frequencies, the large quasiperiodic transient introduces additional noise.
Because the transient is decaying over a rather long time-scale, it is quite narrow in frequency, and does not perturb the measurement spectrum much.

\begin{figure}
\begin{center}
\ieps{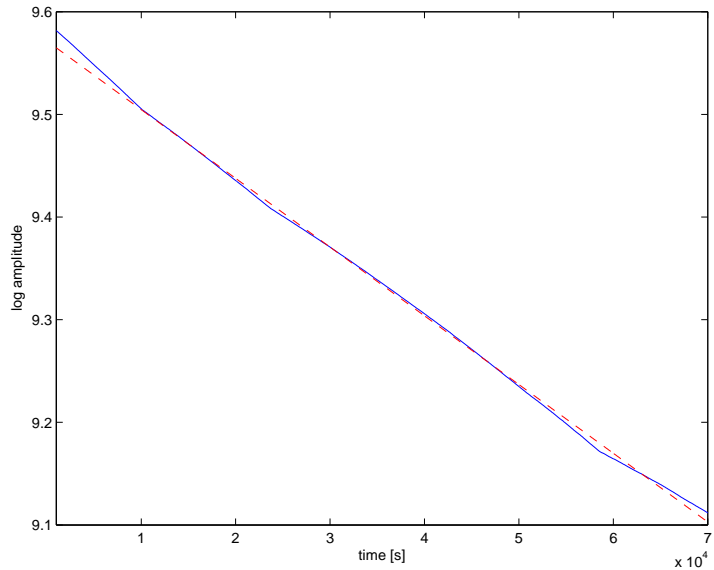}
\end{center}
\caption{The logarithm of the amplitude of the envelope of the closed loop laser noise as a function of time (continuous line), and the best-fitted straight line (dashed line).}
\label{fig:run63_transient}
\end{figure}

Figure \ref{fig:run63_transient} shows as a function of time the envelope of the closed loop laser noise, which is dominated by the quasiperiodic transient.
The fit to an exponential is reasonably good, with a decay time of $1.49 \cdot 10^5$ s.
For comparison, the controller gain at $f = 1/(2 \times 16.7 {\rm s}) = 0.03 {\rm Hz}$ is 72.8 dB, so that the predicted transient decay rate from Eq. \ref{eq:decayrate} is $1.46 \cdot 10^5$ s, in good agreement with what is observed in the simulations.

\section{Time delay interferometry} \label{TimeDelayInterferometry}
From the diagram of Fig. \ref{fig:topology}, we can write:
\begin{equation}
s_{31}(t) = p_{31}(t) + n_{31}(t) - \left\{ 2\delta_{31}(t) - \Delta_{31}(t) + p_{13}(t-L_2[t]) \right\},
\end{equation}
and similarly for the other signals $s_{13}$, $s_{12}$ and $s_{21}$.
Here, $s_{31}$ and $s_{13}$ are the signals measured on S/C 1 and 3, respectively, for the phase difference along arm 2, while $s_{21}$ and $s_{12}$ are the signals on S/C 1 and 2 for the phase difference along arm 3.
On a given S/C, the relative phase between the two lasers is measured by bouncing the light from one laser off the proof mass on its optical bench, and then sending the reflected light through an optical fiber to the other optical bench, where it interferes on a photodiode with the laser from that bench.
The measurement resulting from sending $p_{31}$ to the bench where $p_{21}$ is located, for instance, is given by
\begin{equation}
\tau_{21}(t) = p_{31}(t) - p_{21}(t) - 2\delta_{31}(t) + 2\Delta_{31}(t) + \mu_{21}(t),
 \end{equation}
where $\mu_{21}$ is the noise introduced by the fiber.

Time delay interferometry (TDI) is a set of combinations of the $s_{ij}$ and $\tau_{ij}$ measurements that have the property of cancelling the laser noises ($p_{ij}$) and the S/C acceleration noises ($\Delta_{ij}$) while preserving a good response to gravitational waves.
Successive versions of TDI variables of increasing complexity have been developed for LISA configurations with static arms \cite{TDI1,TDI2,TDI3,TDI4}, and for rotating and stretching arms \cite{TDI5,TDI6,TDI7}.
The latter, more complicated versions are known as ``second generation'' TDI variables, and are necessary for the small errors in the cancellation of the laser noise to be below secondary noises, when the initial laser noise level is that of a stabilized laser without locking, $\sim 30 {\rm Hz}/\sqrt{\rm Hz}$.

\begin{figure}
\begin{center}
\ieps{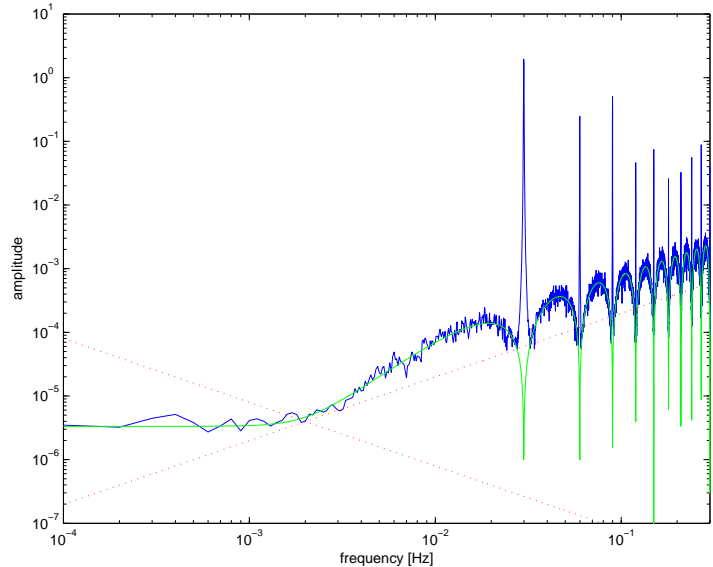}
\end{center}
\caption{(Color online) The amplitude spectrum of the $X$ TDI variable from our simulations (blue), and the theoretical spectrum (green) constructed from the proof mass acceleration noises ($\delta_{ij}$, all with amplitude spectrum as described by the dotted line with a negative slope), and from the optical path noises ($n_{ij}$, all with amplitude spectrum as described by the dotted line with a positive slope). The spectrum was constructed from $7\cdot 10^4$ s of simulated data sampled at 100 Hz. The amplitude units are arbitrary, but are the same as in Fig. \ref{fig:psd}.}
\label{fig:TDIX}
\end{figure}

\begin{figure}
\begin{center}
\ieps{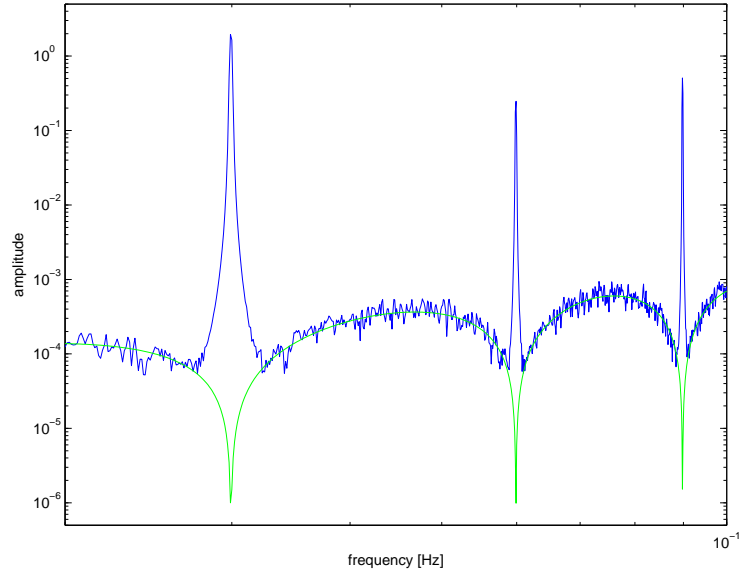}
\end{center}
\caption{(Color online) Close-up of Fig. \ref{fig:TDIX} for $0.02 {\rm Hz} < f < 0.1 {\rm Hz}$.}
\label{fig:TDIXzoom}
\end{figure}

Figures \ref{fig:TDIX} and \ref{fig:TDIXzoom} show the first generation TDI variable $X$ that is formed with self-locking.
For reference, $X$ is constructed in the following manner:
\begin{eqnarray}
X(t) = s_{12}[t-L_3(t)-2 L_2(t)] - s_{13}[t-L_2(t)-2 L_3(t)] + s_{21}[t-2 L_2(t)] - s_{31}[t-2 L_3(t)] \nonumber \\
 + s_{13}[t-L_2(t)] - s_{12}[t-L_3(t)] + s_{31}[t] - s_{21}[t] \nonumber \\
 - \frac{1}{2} \left\{ -\tau_{31}[t-2 L_2(t)-2 L_3(t)] + \tau_{31,}[t-2 L_2(t)] + \tau_{31}[t-2 L_3(t)] - \tau_{31}[t] \right. \nonumber \\
\left. + \tau_{21}[t-2 L_2(t)-2 L_3(t)] - \tau_{21}[t-2 L_2(t)] - \tau_{21}[t-2 L_3(t)] + \tau_{21}[t] \right\},
\end{eqnarray}
where
\begin{equation}
L_2(t) = 16.7 {\rm s} + \frac{10 {\rm m/s}}{c} t 
\end{equation}
and
\begin{equation}
L_3(t) = 16.7 {\rm s} + \frac{1 {\rm m/s}}{c} t .
\end{equation}
It should be noted that the time series $s_{ij}$ and $\tau_{ij}$ used to construct $X$ are only sampled at 100 Hz, and that simple linear interpolation between the two nearest points is used to apply the right time-shift.
In other words, the fact that the data from the simulations and the theoretical model agree very well in Fig. \ref{fig:TDIX} shows that laser self-locking allows the construction of $X$ with rather loose requirements on the accuracy of the time-shifts.
In fact, we have verified directly that adding a fixed delay of 100 $\mu$s to $L_2$ when constructing $X$ does not degrade appreciably the noise spectrum.
With a 1 ms delay added, however, the $X$ amplitude spectrum is increased by a factor of $\sim 10$ at frequencies above 1 mHz.
A fixed delay is used for the whole simulation (a total of $\sim 19.4$ hours) to simulate the case where the arm length measurements are not being performed continuously; this could be the case if, for instance, these measurements perturb the gravitational wave measurements, or if they involve the exchange of information with ground stations.

Without laser locking, the arm lengths have to be known to a much greater accuracy in order to implement TDI.
Ref. \cite{TDIImplementation} shows that the required accuracy is of the order of 100 ns.
We have verified directly by constructing $X$ as described above with open loop simulation data that the simulated spectrum was 3-4 orders of magnitude above the theoretical expectations.
This, of course, is a direct result of our 100 Hz data sampling and two-point interpolation scheme, which are far from being accurate enough to implement TDI with a large laser noise.

\section{Conclusion}
Starting from the idea presented in Ref. \cite{SGMS} to lock the two lasers of a LISA arm to each other in order to improve their phase stability, we have implemented a realistic simulation of the system, which includes a stable control loop design (with unity gain frequency at 10 kHz), significant arm length variations ($\sim 10$ m/s) from the orbital motion of the S/C, and the standard secondary noises (optical path noise, proof mass acceleration noise, and S/C acceleration noise). 
We have shown that a reduction of the amplitude of the laser phase noise by as much as seven orders of magnitude was indeed possible at low frequencies.
We have also shown that the large quasi-periodic transient that is produced by the locking procedure is very stable and decays exponentially on a time-scale of hundreds of thousands of round-trip light-times.
We argue that this is the reason why it only adds noise in narrow frequency bands, so that its presence is inconsequential for the data analysis.

We have also used our simulated data to implement Time Delay Interferometry (TDI).
Our results show that the same level of noise can be achieved as in the baseline design where no locking is used, except in narrow bands at the frequency defined by the inverse of the round-trip light-time, and its harmonics.
Since the laser noise is much smaller when self-locking is used, the accuracy on the arm lengths required to implement TDI can be significantly less stringent than in the baseline design, $\sim 100 \mu$s vs. $\sim 100$ ns.
We also verify that the first generation TDI variable $X$ can bring the laser phase noise below the secondary noises, even in the presence of significant time dependent changes of the arm lengths (``flexing'').

Our results suggest that laser self-locking could be used to move technical complexity and risk from the implementation of second generation TDI to the implementation of the control loops required to lock the lasers. 
It appears that self-locking and the baseline design are mostly equivalent from the point of view of the instrument sensitivity, so that the viability of both schemes should be carefully evaluated from an engineering point of view.
In particular, it should be verified that more detailed orbital models, which might include rotation of the constellation in addition to the stretching of the arms, do not lead to degraded performances for the self-locking scheme.

\acknowledgments
The author thanks J. W. Armstrong, T. A. Prince, B. L. Schumaker and D. A. Shaddock for useful discussions.
This research was performed at the Jet Propulsion Laboratory, California Institute of Technology, under contract with the National Aeronautics and Space Administration.


\begin{references}
\bibitem{Rotating} If the S/C constellation is rotating, the light-times might be different for counter-propagating beams.
\end{references}
\end{document}